\documentclass[prb,10pt,amsmath,amssymb,letterpaper,nobalancelastpage,aps,showpacs,preprintnumbers,twocolumn]{revtex4-1}
\usepackage{graphicx}
\usepackage{dcolumn}
\usepackage{bm}
\usepackage{slashed}
\usepackage{amssymb}
\usepackage{amsmath}
\usepackage{amsthm}

\begin{document}

\title{Bulk-Edge correspondence of entanglement spectrum in 2D spin ground states}

\author{Raul A. Santos}
\email{santos@insti.physics.sunysb.edu} 
\affiliation{C.N. Yang Institute for Theoretical Physics,\\
       Stony Brook University,\\
  Stony Brook, NY 11794-3840, USA}

\begin{abstract}
General local spin $S$ ground states, described by a Valence Bond Solid (VBS) on a two dimensional lattice are studied. 
The norm of these ground states is mapped to a classical $O(3)$ model on the same lattice. Using this quantum-to-classical mapping we obtain 
the partial density matrix $\rho_{A}$ associated with a subsystem ${A}$ of 
the original ground state. We show that the entanglement spectrum of $\rho_{\rm A}$ in a translation invariant lattice 
is given by the spectrum of a quantum spin chain at the boundary of region $A$, with local Heisenberg type interactions between spin $1/2$
particles.
\end{abstract}

\preprint{YITP-SB-12-30}

\maketitle

\section{Introduction}
Quantum entanglement, the spooky action at a distance that has been signaled as \textit{the} characteristic of quantum mechanics \cite{Sch1935}, 
has received renewed attention recently, specially with the growth of quantum information science \cite{Nielsen_Chuang}, and as a new tool 
to study properties of many-body systems \cite{Amico2008,Eisert2010}. 
It has been found that entanglement is sensitive to topologically ordered states \cite{Kitaev2006,Levin2006}, quantum phase transitions \cite{Wei2005},  
and even magnetic properties of solids \cite{Ghosh2003}.

A complete description of the entanglement properties of a bipartite pure state $|\Psi\rangle$ composed of subsystems ${A}$ and ${B}$ is given by 
the entanglement spectrum (ES) \cite{Li2008}, i.e. the eigenvalues of the reduced density matrix (RDM) of subsystem ${A}$ (or ${B}$).
The RDM for the subsystem $A$, $\rho_A$, is obtained by tracing out the degrees of freedom belonging to $B$ from the density matrix $\rho=|\Psi\rangle\langle \Psi|$ 
which characterizes the state $|\Psi\rangle$. In general the density matrix of the subsystem $A$ may be written as $\rho_A=\exp(-\beta H_{\rm eff})$, 
where $H_{\rm eff}$ is an effective (also called entanglement) Hamiltonian.
It has been shown that the entanglement Hamiltonian describes excitations living at the edge of partitions of the ground state
of fractional quantum Hall states \cite{Li2008}, one dimensional \cite{Pollmann2010,Calabrese2008,Santos2012a,Santos2012b,Turner2011,Tomale2010,Orus2005,Tu2011} and topological systems \cite{Swingle2012,Qi2012,Fidkowski2010}. 
In two dimensions, few analytical results on the entanglement spectrum in generic spin systems are known \cite{Dubail2011,Yao2010}. Numerical studies 
of the ES have been performed in the 2D Affleck, Lieb, Kennedy and Tasaki (AKLT) model \cite{Affleck1987}, which possess a known valence bond 
solid (VBS) ground state. 

Lou {\it et al}\cite{Lou2011} and Cirac {\it et al.}\cite{Cirac2011} showed that the ES of a partition in the ground state of the AKLT model is 
related with the conformal {\rm XXX} Heisenberg model on the boundary of the partition by using Montecarlo and projected entangled pair states 
(PEPS) \cite{Verstraete2004} in finite size systems.
In this paper we show that the ES of a partition of a whole class of ground states defined in translational invariant lattices, can be 
approximated by the thermal spectrum of a series of local Hamiltonians which are the conserved charges associated with the {\rm XXX} Hamiltonian 
defined on the boundary of the partition. 

In this paper we introduce the spin $S$ model in section \ref{model}, defined on a two dimensional lattice wrapped on a torus and construct its explicit VBS ground state
following \cite{Affleck1987,Korepin2010}. Then, we derive an expression for the RDM (also called partial density matrix) $\rho_A$  in section \ref{RDMcons}. This operator
is expressed in terms of classical variables in section \ref{mappingQC}. In this representation, the operator can be expanded in different
graph contributions of the classical $O(3)$ model as presented in section \ref{graph1}. From this expression we identify the Heisenberg Hamiltonian for spin $1/2$ particles in the boundary as the 
leading term in a sequence of boundary Hamiltonians. Evidence for the structure of the entanglement Hamiltonian is given in section \ref{O_N} based
on the analysis in the continuous limit. In the last section, we summarize the results and discuss further possible generalizations.

\section{Spin $S$ VBS ground state on a two dimensional torus}\label{model}
\begin{center}
\begin{figure}
\includegraphics[scale=.5]{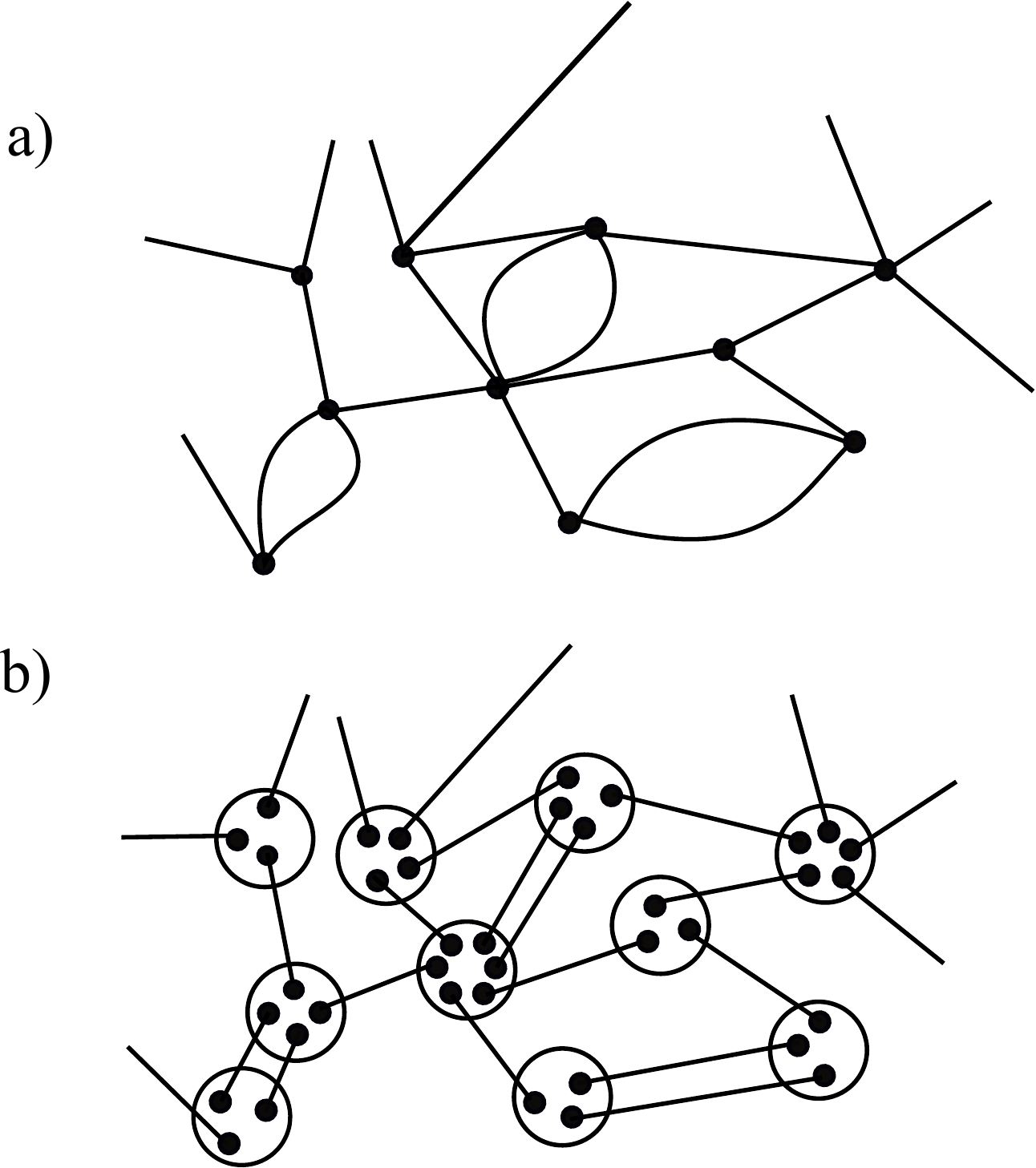}
 \caption{a) Original planar graph $\mathcal G$, vertices represented with black dots. b) VBS state on $\mathcal G'$, circles (vertices of $\mathcal G'$)
represent symmetrization of constituents spin $1/2$ (black dots) particles, while bonds represent anti-symmetrization of neighboring spins.
Note that any loop in $\mathcal G$ would make the associated VBS state vanish, as it would correspond to the antisymmetrization of a state with 
itself.}\label{fig:isomorphism}
\end{figure}
\end{center}

As discussed on \cite{Affleck1987,Katsura2010, Korepin2010} it is possible to construct a valence bond solid (VBS) ground state in a planar 
graph $\mathcal G$ (without edges starting and ending in the same site) 
in the following way: Given a planar graph $\mathcal G$, consisting of a set of vertices (sites) $V$ and edges $E$, with $z_i$ edges arriving 
to vertex $i$ (in graph theoretical language, $z_i$ is called coordination number), we place a local spin $S_i$ on the vertex with the condition $S_i=z_i/2$. The local spin state is constructed from the symmetric subspace of $z_i$ 
spins 1/2 (doing this we obtain a higher spin representation of dimension $2S_i+1$ from $2S_i$ fundamental representations of $SU(2)$). Finally 
we antisymmetrize between nearest neighbors. Representing the spin $1/2$ constituents of the spin $S_i$ at site $i$ as black dots, using a 
circle to indicate symmetrization and a bond between antisymmetric neighbors, we obtain a planar graph $\mathcal G'$ isomorphic to $\mathcal G$, see Fig \ref{fig:isomorphism}.

The AKLT Hamiltonian for which the VBS state constructed is a ground state is a sum over interactions on all edges $E$ of $\mathcal G$,
$H=\sum_{\langle k,l\rangle\in E} H_{kl}(\vec{S}_k+\vec{S}_l),$ where the Hamiltonian density $H_{kl}$ is 
\begin{equation}\label{Hdensity}
H_{kl}(\vec{S}_k+\vec{S}_l)=\sum_{J=S_k+S_l+1-M_{lk}}^{S_k+S_l}A^J_{kl}\pi^J_{kl}(\vec{S}_k+\vec{S}_l),
\end{equation}
\noindent  the coefficients $A^J_{kl}>0$ are arbitrary and can depend on the edge $\langle k,l\rangle$, while the positive number $M_{kl}$
is the number of bonds (edges) connecting the sites $k$ and $l$. The operator $\pi^J_{kl}(\vec{S}_k+\vec{S}_l)$
is a projector of the total spin $\vec{J}_{kl}=\vec{S}_l+\vec{S}_k$ of the edge $\langle k,l\rangle$ on the subspace of spin 
value $J$, its explicit form is

\begin{eqnarray}
 \pi^J_{kl}(\vec{J}_{kl})=\prod_{j=|S_k-S_l|,j\neq J}^{S_k+S_l}\frac{(\vec{J}_{kl})^2-j(j+1)}{J_{kl}(J_{kl}+1)-j(j+1)},
\end{eqnarray}

The VBS state is the {\it unique} ground state of $H$ \cite{Kirillov1989}.
While this construction is totally general, in the rest of this discussion we will focus on graphs without boundaries, 
which can be embedded on a two dimensional torus, with $M_{ij}=1$ for all edges.

\section{Partial density matrix and Schwinger boson representation of {\rm VBS} ground state}\label{RDMcons}

In this section, we introduce a general way of writing the reduced density matrix of a pure system in terms of overlap matrices. These
matrices have elements which correspond to overlap amplitudes between states spanning the ground space of Hamiltonians defined entirely in 
the subsystems. We apply these results to the VBS case introduced in the previous section.

Using the Schmidt decomposition, any ground state $|\Psi\rangle$ of a system can be written as 

\begin{equation}\label{decomp}
|\Psi\rangle=\sum_{\alpha}|A_\alpha\rangle\otimes|B_\alpha\rangle, 
\end{equation}

\noindent where the states $|A_\alpha\rangle$ and $|B_\alpha\rangle$
are related to the usual states appearing in the Schmidt decomposition by a scale factor. $|A_\alpha\rangle$ and $|B_\alpha\rangle$ are 
states defined in the subsystems $A$ and $B$, with associated Hilbert spaces $\mathcal{H}_A$ and $\mathcal{H}_B$ respectively. The total 
system has a Hilbert space $\mathcal{H}=\mathcal{H}_A\cup\mathcal{H}_B$.  The set of states $\{|A_\alpha\rangle,|B_\alpha\rangle\}$ is a complete, 
linear independent but not orthonormal basis (in principle). The density matrix for this pure state is the projector onto the ground state $\rho={\mathcal N}|\Psi\rangle\langle \Psi|$, with 
$\mathcal N^{-1}=\langle \Psi|\Psi\rangle$. Tracing out the sites
belonging to the subsystem $B$, we obtain the partial density matrix, which describe the system $A$, $\rho_A={\rm Tr}_B\rho$. 
Using (\ref{decomp})
the partial density matrix becomes $\rho_A={\mathcal N}\sum_{\alpha\beta}\langle B_{\beta}|B_\alpha\rangle|A_\alpha\rangle\otimes\langle A_\beta|$. 
Using standard algebraic techniques \cite{com1}, the partial density matrix can be written as

\begin{equation}\label{density_M}
 (\rho_A)_{\mu\alpha}=\sum_{\gamma}\frac{\langle (A_{\mu}|A_\gamma\rangle)^*\langle B_{\gamma}|B_\alpha\rangle}{\langle\Psi|\Psi\rangle}.
\end{equation}

From the Schmidt decomposition, we know that the dimension of this operator is the minimum between the dimensions of $\mathcal{H}_A$ and 
$\mathcal{H}_B$. Let's assume $\dim\mathcal{H}_A\leq\dim\mathcal{H}_B$. The dimension of $\rho_A$ is then $\dim\mathcal{H}_A\times\dim\mathcal{H}_A$.
This matrix is not hermitian in the usual sense $O^\dagger=O$, but it is isospectral with $(\rho_A)^\dagger$.

\begin{center}
\begin{figure}
\includegraphics[scale=.7]{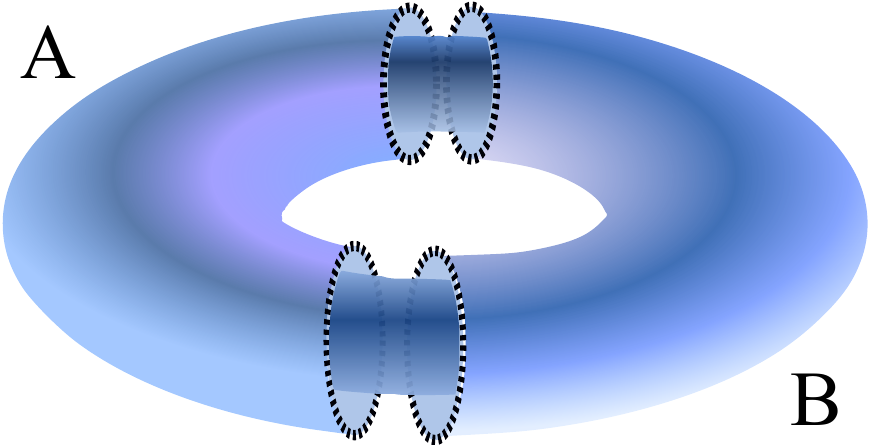}
 \caption{(color online) a.- The VBS ground state in its tensor product representation can be viewed as a two 
dimensional lattice build up from contractions of virtual indices (black lines in the plane). The physical indices stick out of the plane. 
After making the partition, virtual indices at the boundary are free. b.- The overlap matrix $M^{[A]}_{\alpha\beta}=\langle A_\alpha|A_\beta\rangle$ 
can be obtained by stacking two of this systems and contracting their physical index \cite{Santos2012a}. This creates a two layer stack.
c.- Graphical representation of the partial density matrix $\rho_A$, for a particular partition. 
For periodic boundary conditions, the overlap matrix corresponds to a section of the torus with two different, inner and outer, layers.
To compute $(\rho_A)_{\alpha\beta}$, we glue the inner layers of the overlap matrices $M^{[A]}$ and $M^{[B]}$ (contracting the virtual indices),
 obtaining a two layer torus with a cut in the outer sheet along the boundary of the partition. The cut here is represented by the dashed line.}
\label{fig:torus}
\end{figure}
\end{center}

Using the Schwinger boson representation for spin operators, the VBS state on  $\mathcal G$ can be written as \cite{Arovas1988, Kirillov1989} 

\begin{equation}
 |\Psi_{vbs}\rangle=\prod_{\langle i ,j \rangle \in E_{\mathcal G}}(a^\dagger_ib^\dagger_j-a^\dagger_jb^\dagger_i)|0\rangle,
\end{equation}

\noindent where $E_{\mathcal G}$ is the set of all edges (bonds) of $\mathcal G$ and $|{0}\rangle$ is the state annihilated by all the
$a_i$ and $b_i$ operators, i.e. $a_i|0\rangle=b_i|0\rangle=0$, $\forall$ $i$. 
For a generic partition of the system into two subsystems $A$ and $B$ (we assume both of them to be connected regions) with boundaries $
\partial A$ and $\partial B$, we have a collection of 
vertices $V_A$, $V_B$ such that $V_A \cup V_B = V_{\mathcal G}$ and a collection of edges (bonds) which endpoints live either both in $A$ 
($B$) or one in $A$ and the other in $B$. For bonds which both endpoints live in $A$ we will say $\langle i,j \rangle\in E_A$ (similarly for $B$),
while for shared bonds with endpoints $i$ and $j$ we use $i\in\partial A, j\in \partial B$.
The set of shared bonds we will call it $\partial$ (and is the same for $A$ and $B$). Finally the cardinality of a set $M$ is denoted by $|M|$.
Using this definitions, we can write the state 
$|\Psi_{vbs}\rangle$  in the form (\ref{decomp}) as follows; first we write

\begin{eqnarray}\label{split}
 |\Psi_{vbs}\rangle=\prod_{\langle i,j\rangle \in \atop E_{A}\cup E_{B}}(a^\dagger_ib^\dagger_j-a^\dagger_jb^\dagger_i)\prod_{ i\in \partial A \atop j \in\partial B}(a^\dagger_ia^\dagger_j+b^\dagger_jb^\dagger_i)|0\rangle,
\end{eqnarray}

\noindent where we have applied a local basis transformation on the sites (vertices) in $B$, $a^\dagger_i\rightarrow -b^\dagger_i$ and 
$b^\dagger_j\rightarrow a^\dagger_j$ just for later convenience. 
In the shared bonds, we can assign to an endpoint $j$ of a bond, it's partner in the other end of the bond to be $\bar{j}$. Doing this we can 
expand (\ref{split}) in the form \cite{Katsura2010}

\begin{eqnarray}\label{split2}\nonumber
 |\Psi_{vbs}\rangle&=&\sum_{\{\alpha\}}\prod_{ i\in \partial }(a^\dagger_i)^{\alpha_i}(b^\dagger_i)^{1-\alpha_i}(a^\dagger_{\bar{i}})^{\alpha_i}(b^\dagger_{\bar{i}})^{1-\alpha_i}\\\nonumber
&\times&\prod_{\langle i,j\rangle \in E_{A}\cup E_{B}}(a^\dagger_ib^\dagger_j-a^\dagger_jb^\dagger_i)|0\rangle\\
&=&\sum_{\{\alpha\}}|A_{\{\alpha\}}\rangle\otimes|B_{\{\alpha\}}\rangle,
\end{eqnarray}

\noindent here $\{\alpha\}=\{\alpha_1,\alpha_2,..\alpha_{|\partial|}\}$ with $\alpha_i=0,1$, labels the different ground states of the subsystems, 
which span a Hilbert space of dimension $2^{|\partial|}$. The Hamiltonian in subsystem $A$ is defined by $H_A\equiv\sum_{\langle k,l\rangle\in E_A}H_{kl}$ (and similarly
for $B$), with $H_{kl}$ given by (\ref{Hdensity}). From (\ref{split2}), we can read off the form of the states $|A_{\{\alpha\}}\rangle$

\begin{equation}\label{bulk_state}
 |A_{\{\alpha\}}\rangle=\prod_{\langle i,j\rangle \in E_{A}}(a^\dagger_ib^\dagger_j-a^\dagger_jb^\dagger_i)\prod_{ i\in \partial }(b^\dagger_i)^{1-\alpha_i}(a^\dagger_i)^{\alpha_i}|0\rangle,
\end{equation}

\noindent using (\ref{density_M}) and (\ref{bulk_state}), we can compute the density matrix $\rho_A$ in terms of the overlap matrices 
$M^{[A]}_{\{\alpha\},\{\beta\}}=\langle A_{\{\alpha\}}|A_{\{\beta\}}\rangle$. From eq. (\ref{density_M}), the partial density matrix is constructed
 gluing together two of these overlap matrices, one for each subsystem, along the boundary of the partition, leaving one index free in each overlap matrix, 
obtaining a torus with a cut along the partition (see Fig. \ref{fig:torus}). 

From this construction, we see that we can write $\rho_A$ as a block diagonal
operator, with a nontrivial block of dimension $2^{|\partial|}\times 2^{|\partial|}$, and a trivial block (full of zeros), of dimension
$(\dim\mathcal{H}_A-2^{|\partial|})\times(\dim\mathcal{H}_A - 2^{|\partial|})$. This result can be understood from the properties of the 
VBS state. This state is annihilated at each and every site by the action of the Hamiltonian density $H_{kl}$. After making the partition
the states defined in the subsystems are still annihilated by the local Hamiltonians defined in each partition, but the states of the 
sites at the edges who cross from one subsystem to the other (in our notation, the edges belonging to the set $\partial$) are free to
have any possible state on them, as no local Hamiltonian defined in just one subsystem can act on this edges. This feature has been
encountered before in the study of AKLT chains, where the dimension of the partial density matrix does not increase with the size of the 
system \cite{Fan2004}.

The computation of this overlap matrix can be mapped to the computation
of partition and correlation functions in an $O(3)$ model, by means of the classical representation of the VBS state \cite{Arovas1988},
as we show in the next section.

\section{Quantum to Classical mapping}\label{mappingQC}

Introducing the spinor coordinates $\phi^a_k=(u_k,v_k)=(e^{i\varphi_k/2}\cos\frac{\theta_k}{2},e^{-i\varphi_k/2}\sin\frac{\theta_k}{2})$
at site $k$, with $\theta\in[0,2\pi]$, $\varphi_k\in[0,2\pi)$, we can define the the spin coherent state $|\Omega_k\rangle$ as

\begin{equation}
|\Omega_k\rangle=\frac{(u_ka^\dagger_k+v_kb^\dagger_k)^{2S_k}}{\sqrt{(2S_k)!}}|0_k\rangle,
\end{equation}

\noindent ($|0_k\rangle$ being the vacuum state at site $k$), this states are complete but not orthogonal. Inserting the resolution of the identity 

\begin{equation}
 1_{2S_k+1}=\frac{2S_k+1}{4\pi}\int d\Omega_k|\Omega_k\rangle\langle\Omega_k|,
\end{equation}

\noindent in $M^{[A]}_{\{\alpha\}\{\beta\}}$, and using the result $\langle0|a_k^{S_k-l}b_k^{S_k+l}|\Omega_k\rangle=\sqrt{(2S_k)!}u_k^{S_k-l}v_k^{S_k+l}$, 
 the following form of the overlap matrix is obtained (dropping overall constant factors)

\begin{eqnarray}\label{ClM}
 M^{[A]}_{\{\alpha\}\{\beta\}}&=&\int\prod_{i\in A}\frac{d\Omega_i}{4\pi}\prod_{\langle i,j\rangle\in E_A}(1-\hat{\Omega}_i\cdot\hat{\Omega}_j)\\\nonumber
&\times&\prod_{k\in\partial A}(u_k)^{\alpha_{k}}(v_k)^{1-\alpha_{k}}(u_k^*)^{\beta_{k}}(v^*_k)^{1-\beta_{k}},
\end{eqnarray}

\noindent here $\hat{\Omega}_k=(\sin\theta_k\cos\varphi_k,\sin\theta_k\sin\varphi_k,\cos\theta_k)$ is the unit vector over the two dimensional
sphere $S^2$ and $u^*$ is the complex conjugate of $u$. From (\ref{ClM}) we see that the overlap matrix $M^{[A]}$ is hermitian, so the partial
density matrix $\rho_A={\mathcal N}(M^{[A]})^*M^{[B]}$ is also hermitian. Using now that $(u_k)^{1-\alpha_{k}}(v_k)^{\alpha_{k}}=\phi_k^{\alpha_k}$ (abusing notation, $\alpha_k$ goes from being a power, to
become a (supra)index, $\phi_k^0=u_k$, $\phi_k^1=v_k$) and the identity

\begin{equation}
 2\phi_k^\alpha(\phi_k^*)^\beta=\delta_{\alpha\beta}+\hat{\Omega}_k\cdot\vec{\sigma}_{\alpha\beta},
\end{equation}

\noindent where $\delta_{\alpha\beta}$ is the Kronecker delta symbol, and $\vec{\sigma}=(\sigma_1,\sigma_2,\sigma_3)$ is a vector of
Pauli matrices (no distinction is made between upper or lower greek indices); the expression for the overlap matrix can be written as

\begin{eqnarray}
 M^{[A]}_{\{\alpha\}\{\beta\}}=\int\prod_{i\in A}\frac{d\Omega_i}{4\pi}\prod_{\langle i,j\rangle\in E_A}(1-\hat{\Omega}_i\cdot\hat{\Omega}_j)\\\nonumber
\times\prod_{k\in\partial A}({\mathbb I}+\hat{\Omega}_k\cdot\vec{\sigma})_{\alpha_k\beta_k},
\end{eqnarray}

\noindent combining this result with (\ref{density_M}), the density matrix of the subsystem $A$ becomes

\begin{flalign}\label{DM_Cl}
(\rho_A)_{\{\alpha\}\{\beta\}}=&\frac{1}{Z}\int\prod_{k\in \mathcal G}\frac{d\Omega_k}{4\pi}\prod_{\langle i,j\rangle\in E_A\cup E_B}(1-\hat{\Omega}_i\cdot\hat{\Omega}_j)\\\nonumber
\times&\prod_{\langle k,l\rangle\in\partial}[({\mathbb I}+\hat{\Omega}_k\cdot\vec{\sigma})({\mathbb I}+\hat{\Omega}_{l}\cdot\vec{\sigma})]_{\alpha_k\beta_l}.
\end{flalign}

\noindent with $Z$ the proper normalization factor to make ${\rm Tr}{\rho_A}=1$. We can expand the matrix product inside (\ref{DM_Cl}) using the 
product identity for Pauli matrices $\sigma_i\sigma_j=\delta_{ij}{\mathbb I}+i\epsilon_{ijk}\sigma_k$ (repeated index implies sum)
where $\epsilon_{ijk}$ is the totally antisymmetric Levi-Civita tensor. The result of the term inside the square bracket in (\ref{DM_Cl}) is then

\begin{flalign}\label{2bonds}
&[({\mathbb I}+\hat{\Omega}_k\cdot\vec{\sigma})({\mathbb I}+\hat{\Omega}_{l}\cdot\vec{\sigma})]_{\alpha\beta}\\\nonumber
&=(1+\hat{\Omega}_k\cdot\hat{\Omega}_{l})\delta_{\alpha\beta}+(\hat{\Omega}_k+\hat{\Omega}_{l}+i(\hat{\Omega}_k\times\hat{\Omega}_{l}))\cdot\vec{\sigma}_{\alpha\beta},
\end{flalign}

\noindent where $\hat{a}\times\hat{b}$ represent the cross product between vectors $\hat{a}$ and $\hat{b}$.

\section{Graph expansion of the density matrix}\label{graph1}

In this section we derive the structure of the entanglement Hamiltonian as a sequence of spin 1/2 Hamiltonians with increasing
interaction length, using the quantum to classical correspondence introduced in the previous section.

From (\ref{2bonds}), two types of expressions can be assigned to each edge on ${\partial}$. We draw an straight line between $k$ and $l$
whenever in that bond we have the expression $(1+\hat{\Omega}_k\cdot\hat{\Omega}_{l})\delta_{\alpha\beta}$, while we put a wiggly line for
$(\hat{\Omega}_k+\hat{\Omega}_{l}+i(\hat{\Omega}_k\times\hat{\Omega}_{l}))\cdot\vec{\sigma}_{\alpha\beta}.$
Expanding the product over the boundary in (\ref{DM_Cl}), we obtain a sum where each term has 
either a wiggly or straight line corresponding to  $\langle k,l \rangle\in \partial$. All the other bonds who don't belong to $\partial$
have an straight line associated with them.

In general, for a planar graph $\mathcal L$, the expression 

\begin{equation}\label{pf_On}
Z_{O(N)}=\int\prod_{k\in {\mathcal L}}\frac{d\Omega_k}{S_N}\prod_{\langle i,j\rangle\in E_{\mathcal L}}(1+x\hat{\Omega}_i\cdot\hat{\Omega}_j)
\end{equation}

\noindent where $\hat{\Omega}$ is an $N$ dimensional unit vector; corresponds to the partition function over $\mathcal L$ of the $O(N)$ model 
\cite{Nienhuis1987} which is analogous to a model of overlapping loops. To see this, we use that 

\begin{equation}\label{integration}
 \int\frac{d\Omega_k}{S_N}\hat{\Omega}_k\cdot\hat{\Omega}_k=1, \quad  \int\frac{d\Omega_k}{S_N}\prod_{i=1}^{odd}(\hat{\Omega}_i\cdot\hat{\Omega}_k)=0,
\end{equation}

\noindent where $S_N$ is the area of the $S^{N-1}$ sphere and the second property follows form the invariance of the integration measure under change $\hat{\Omega}_k\rightarrow-\hat{\Omega}_k$. 
As the only terms that contribute to $Z_{O(N)}$ are the ones with a product of even $(\hat{\Omega}_i\cdot\hat{\Omega}_k)$ terms at each site of the graph,
the whole partition function can be written as \cite{Nienhuis1982}

\begin{equation}
Z_{O(N)}=\sum_{\mathcal C}w(\zeta,\mathcal C)x^{\Gamma(\mathcal C)}
\end{equation}

\noindent with $\mathcal C$ a particular configuration of loops of total length $\Gamma(\mathcal C)$ that can be embedded in the graph 
$\mathcal L$, and $w(l,\mathcal C)$ being the corresponding weight associated with a loop $\zeta$ and with the particular configuration of 
loops $\mathcal C$. For example, for the hexagonal lattice (coordination number $z_i=3$) each site has associated just two bonds, and each
integration of a site gives a factor of $\frac{1}{N}$, except the last integration which closes the loop. The partition function is then
$Z_{O(N)}=\sum_{\mathcal C}\left(\frac{x}{N}\right)^{\Gamma(\mathcal C)}N^{n(\mathcal C)} $, with $n(\mathcal C)$ the number of loops in the
configuration $\mathcal C$.
The computation of spin correlations $\langle\hat{\Omega}_m\cdot\hat{\Omega}_k \rangle$ corresponds then to the computation of $Z_{O(N)}$,
with configurations that allow loops and open paths that begin at site $m$ and end at site $k$.  
From (\ref{DM_Cl}) expanding the product over the partition's boundary we get a sum over different configurations of loops and open strands in 
the $O(N)$ model, over the graph $\mathcal L$ with defects (wiggly lines). In the present case, $x=-1$ and $N=3$ for the classical partition 
function of the VBS ground state.

\begin{center}
\begin{figure}
\includegraphics[scale=.65]{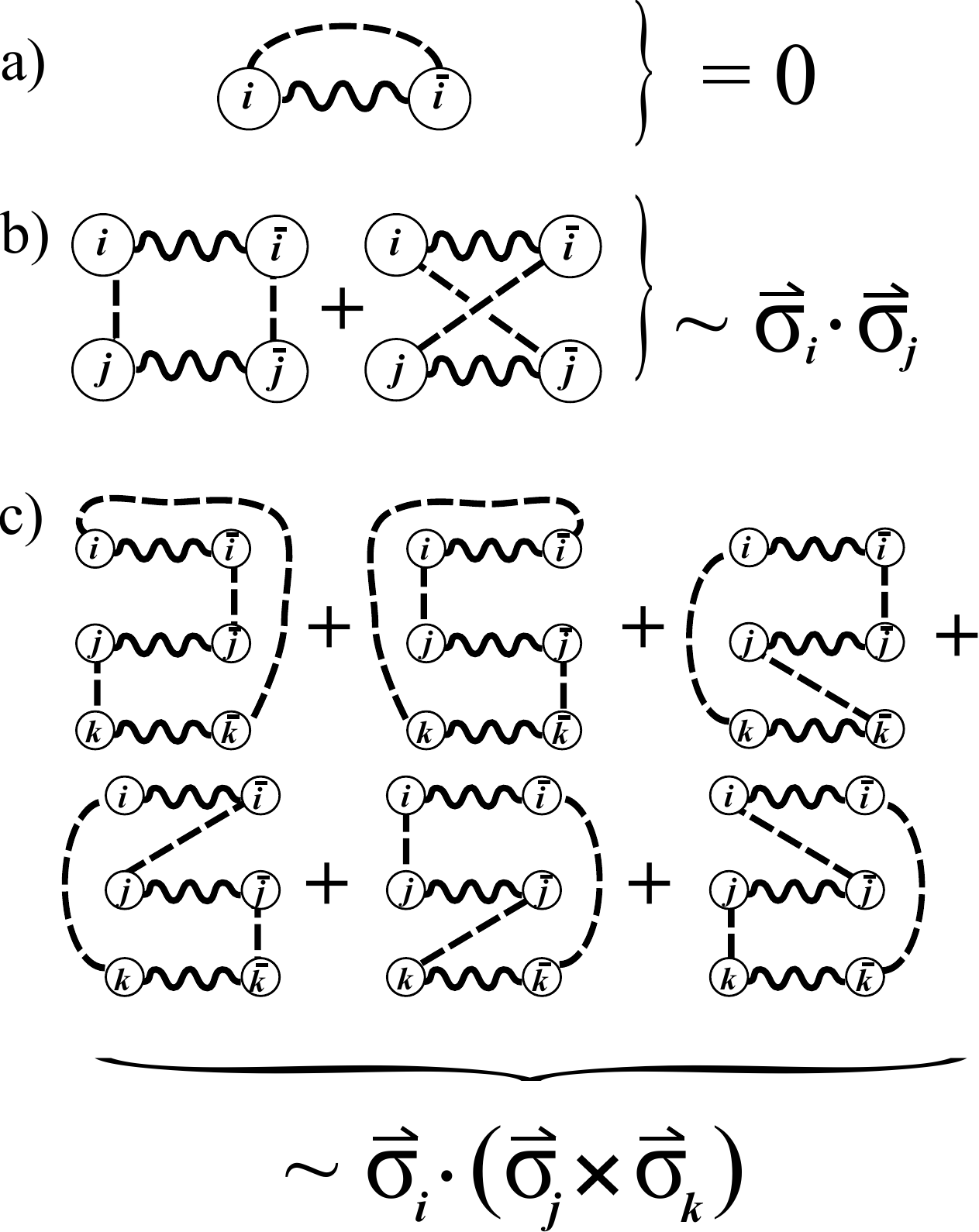}
 \caption{First terms in the graph expansion of the RDM $\rho_A$. For a generic lattice, the number of bonds arriving
to a boundary vertex (big circles) can be any even integer, here we show for simplicity the case corresponding to an hexagonal lattice
where the number of bonds arriving to a vertex is exactly 2. Dashed lines show the remaining bonds after taking the trace over the whole
lattice, except for the boundary vertices joined by wiggly lines.}
\label{fig:graph}
\end{figure}
\end{center}

So far we have developed our ideas for general planar graphs with no loops and no more than one 
bond shared between neighbors ($M_{ij}=1$), but from now on we will focus the discussion on translation invariant lattices with the previous
restrictions. The discussion will remain general for lattices subject to the mentioned restrictions, that can be embedded on a torus. 
Using translation symmetry, we can expand the product over the boundary in (\ref{DM_Cl}) in different contributions of translational invariant
Hamiltonians along the boundary, with increasing number of non-trivial operators (Pauli matrices) acting on the local Hilbert space associated 
with a bond. The first term of the expansion correspond to the identity in the $2^{|\partial|}$-dimensional Hilbert space of the boundary. The
second term, which is proportional to a constant external magnetic field acting on the boundary chain, vanish. This follows from the observation
that in this term, we have just one wiggly bond placed in the boundary - let's say at bond $k$ with endpoints $k$ and $\bar{k}$ - and the rest are just 
straight lines, which after integration will generate all the configurations of loops, {\it and} open lines that start at $k$, travel through
the lattice and end at site $\bar{k}$ (for this type of bonds we will use dashed lines, to indicate the corresponding connection on the lattice). So we 
will have a term which is proportional to the spin correlation between $k$ and $\bar{k}$, and an integral of the form (see fig \ref{fig:graph}.a)

\begin{eqnarray}\nonumber
 \int\frac{d\Omega_k}{4\pi}\frac{d\Omega_{\bar{k}}}{4\pi}(\hat{\Omega}_i\cdot\hat{\Omega}_{\bar{k}})^m(\hat{\Omega}_k+\hat{\Omega}_{\bar{k}}+i(\hat{\Omega}_k\times\hat{\Omega}_{\bar{k}}))\cdot\vec{\sigma}_{\alpha_k\beta_{\bar{k}}},
\end{eqnarray}

\noindent with $m$ odd, which vanish trivially. The next terms in the expansion have two Pauli matrices acting on the different bonds. These terms
are proportional to the only $SU(2)$ invariants that can be constructed with two vectors (of Pauli matrices), namely $\vec{\sigma_i}\cdot\vec{\sigma_j}$
(see fig \ref{fig:graph}.b). Depending on the separation between the wiggly bonds along the boundary, we have different contributions for which
the numerical factor should decay exponentially with this distance, given that the VBS model is expected to have a mass gap (fact that is proven for linear and 
hexagonal lattices \cite{Affleck1987}), result which is in agreement with the $O(N)$ model being noncritical for $N>2$ at $x=-1$ \cite{Guo2000}. 

With the previous results, we can write the following expansion for the density matrix $\rho_A$

\begin{equation}\label{expansion}
\rho_A=\frac{{\mathbb I}}{2^\partial}+\sum_{r,i}A_r\vec{\sigma}_i\cdot\vec{\sigma}_{i+r}+\sum_{ijk}A_{ijk}\vec{\sigma}_i\cdot(\vec{\sigma}_j\times\vec{\sigma}_k)+\dots
\end{equation}

\noindent where ${\mathbb I}$ is the $2^\partial\times 2^\partial$ identity operator and the coefficients $A_r$ and $A_{ijk}$ are related to the correlation functions of the $O(N)$ on the 
lattice $\mathcal L$, with some sites and bonds {\it erased} along the boundary. Specifically for the first coefficient $A_r$ we have

\begin{equation}\label{correlation}
 A_r\sim\langle(\hat{\Omega}_k\cdot\hat{\Omega}_{k+r})(\hat{\Omega}_{\bar{k}}\cdot\hat{\Omega}_{\bar{k}+r})\rangle_{\mathcal L_k}-
\langle(\hat{\Omega}_k\cdot\hat{\Omega}_{\bar{k}+r})(\hat{\Omega}_{\bar{k}}\cdot\hat{\Omega}_{k+r})\rangle_{\mathcal L_k}.
\end{equation}

\noindent Here the correlation function is computed over the lattice $\mathcal L_k$ which is the same lattice $\mathcal L$ but with
the bonds $\langle k, \bar{k}\rangle$ and $\langle k+r, \bar{k}+r\rangle$ erased. This relation is exact for hexagonal lattices, while for 
other lattices with coordination number greater than $3$, all the other possible contractions between even number of legs at the boundary sites
have to be included.  As usual with gapped systems, we expect that
this correlation decays exponentially with the separation of the spins, then we have $A_r\sim\exp(-r/\xi_1)$. Numerical studies for two-leg 
VBS ladders have been performed \cite{Cirac2011} being in agreement with this general result. For $r=1$, the second term in
(\ref{correlation}) vanishes in the thermodynamic limit when minimum distance paths joining the sites are cycles who travel the lattice. Also
taking the limit of infinite size of the $A$ and $B$ subsystems, the interaction between the two boundary chains along different cuts of the 
partition vanishes. Then the total density matrix is the tensor product of matrices with the expansion (\ref{expansion}), for each cut.

\begin{center}
\begin{figure}
\includegraphics[scale=.45]{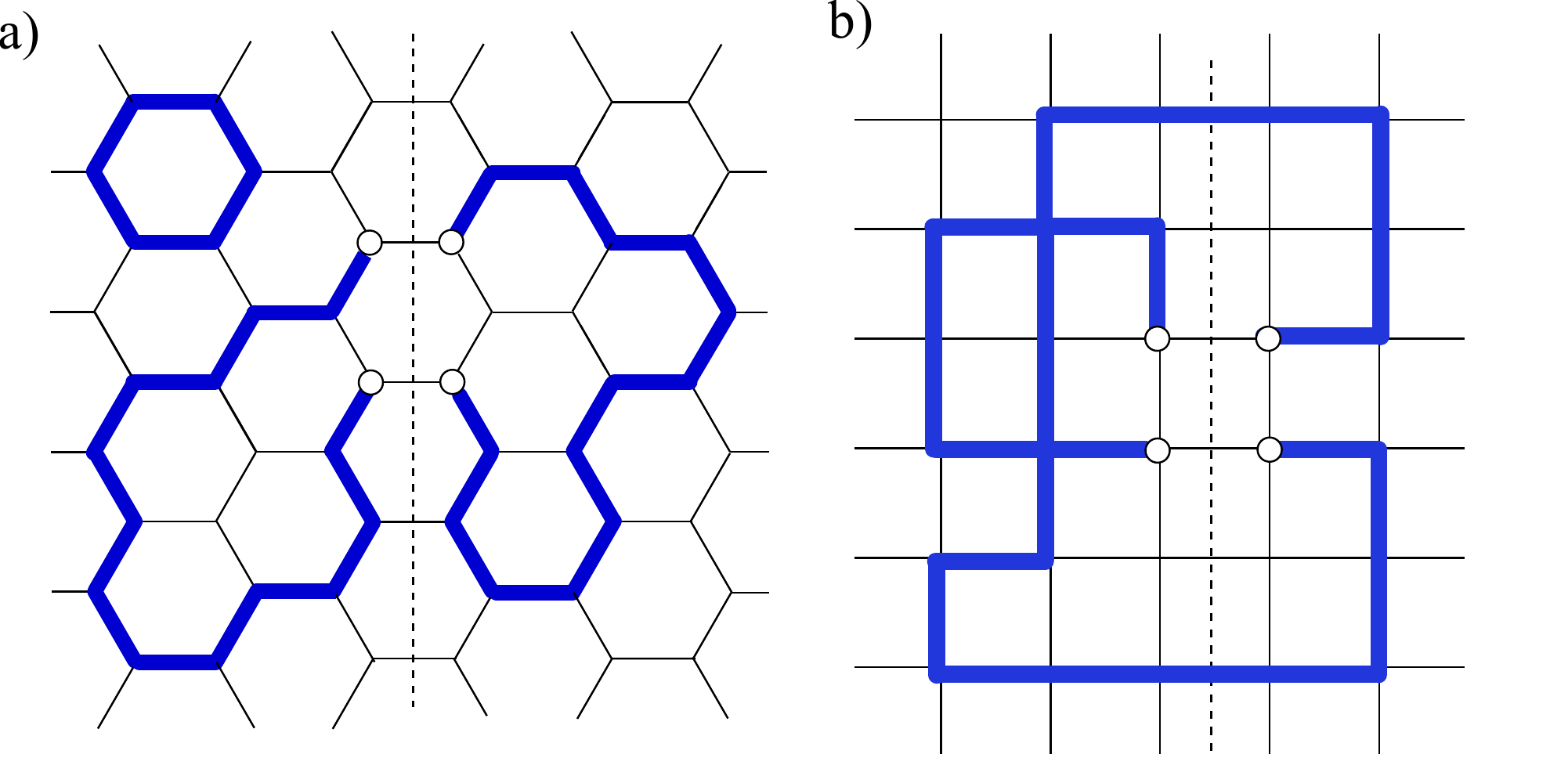}
 \caption{(color online) a. Loop contribution to $A_1$ in the hexagonal lattice. Big circles represent boundary sites, along the partition 
(dashed line). b. Dashed line represent a partition in the square lattice. The contributions to $A_1$ consist now of configurations of overlapping 
loops.}
\label{fig:partition}
\end{figure}
\end{center}

It is clear that the first nontrivial term in the expansion (\ref{expansion}) is the ${\rm XXX}$ Heisenberg Hamiltonian. We can also
determine whether this interaction is ferro or anti-ferromagnetic in the simplest hexagonal lattice, from the loop expansion. The structure 
of the lattice determines the sign of the interaction through the number of bonds that define the allowed paths between site $k$ 
and site $k+1$ (each bond has an associated $x=-1$). An overall minus sign comes from the contraction of two wiggly lines. Then it
is easy to show that for the hexagonal lattice with a partition like the one in Fig. \ref{fig:partition}.a, all the paths connecting the 
boundary sites have even number of bonds, then the sign of the boundary ${\rm XXX}$ Hamiltonian is $-1$, so the boundary chain interaction 
is ferromagnetic. Numerical results \cite{Cirac2011,Lou2011} in finite size square lattices for a partition like Fig \ref{fig:partition}.b, 
indicate that in the square grid the interaction is anti-ferromagnetic.

\section{Continuous limit and Entanglement Hamiltonian}\label{O_N}

In order to unveil the structure of the entanglement Hamiltonian, we can analyze the partial density matrix (\ref{DM_Cl}) taking the lattice
spacing in the original discrete model to zero, obtaining a continuous version of the model. In this limit we can show the locality of the 
boundary (entanglement) Hamiltonian as presented in this section.

Using the identity $\mathbb I +\vec{\sigma}\cdot\hat{\Omega}=\sqrt{2}e^{\frac{i\pi}{4}\vec{\sigma}\cdot\hat{\Omega}},$ we can write 
the product along the boundary of the region $A$ in (\ref{DM_Cl}) as 

\begin{equation}
\prod_{\langle k,l\rangle\in\partial}[({\mathbb I}+\hat{\Omega}_k\cdot\vec{\sigma})({\mathbb I}+\hat{\Omega}_{l}\cdot\vec{\sigma})]
=e^{{\frac{i\pi}{4}\sum(\hat{\Omega}_k+\hat{\Omega}_l)\cdot\vec{\sigma}}},
\end{equation}

\noindent where the sum in the left term runs over the boundary of the subsystem $A$. Putting this result back on (\ref{DM_Cl}), we obtain a 
generating function of a $O(3)$ model with a discrete action $\sum_{\langle i,j \rangle}\ln(1-\hat\Omega_i\cdot\hat\Omega_j)$ and a spin 1/2 
operator localized in the boundary which acts as the current for the generating function. We can study the related $O(N)$ symmetric model with
action $-\sum_{\langle i,j \rangle}\hat\Omega_i\cdot\hat\Omega_j$ which is in the same universality class as (\ref{pf_On}). In this case, the
partial density matrix reads

\begin{flalign}\nonumber
\rho_A[\sigma]=\frac{1}{Z}\int\prod_{k\in \mathcal G}\frac{d\Omega_k}{4\pi}
e^{-\sum_{\langle i,j \rangle}\hat\Omega_i\cdot\hat\Omega_j+\frac{i\pi}{4}\sum_{\langle k,l\rangle\in\partial}(\hat{\Omega}_k+\hat{\Omega}_l)\cdot\vec{\sigma}}
\end{flalign}

\noindent where $\hat\Omega$ is constrained to be a unit vector. In the continuous limit, the reduced density matrix becomes the 
generating functional of $O(3)$ nonlinear sigma model in Euclidean two dimensional space, with an external current localized at the boundary of 
$A$. This Euclidean non-linear sigma model has been well studied \cite{Baarden1976,Brezin1976} and can be solved by standard methods \cite{Peskin}. Here we recall these 
methods for completeness. 

We can impose the unit vector constraint on $\hat\Omega$ by introducing an auxiliary field $\alpha$. In sum, we have

\begin{flalign}\label{DM_Cont}
\rho_A[\sigma]=\frac{1}{Z}\int\mathcal D\Omega \mathcal D \alpha \exp{\left[-S[\Omega,\alpha]-\frac{i\pi}{2}\int d^2 x\,{\vec\Omega}\cdot\vec{\sigma}\right]}
\end{flalign}

\noindent with $\sigma^k(x_1,x_2)=\sigma^k(x_1)(\delta(x_2)+\delta(x_2-L_A))$, $(k=1..3)$ an spin $1/2$ field defined at the boundary of $A$, 
which we have placed conveniently at $x_2=0$ and $x_2=L_A$. The action $S[\Omega,\alpha]$ is given by

\begin{equation}
S[\Omega,\alpha]=\frac{1}{2g^2_0}\int d^2 x\left\{(\nabla\vec\Omega)^2+i\alpha(x)(\vec\Omega(x)^2-1)\right\},
\end{equation}

\noindent where we have introduced a bare coupling $g_0$. As the discussion is essentially the same for any number of components of the
$\Omega$ field, we now consider the more general $N$ component case with the corresponding $O(N)$ global symmetry. We can integrate out the 
field $\Omega$, as the action in this field is quadratic, obtaining

\begin{widetext}

\begin{flalign}\label{Z_Cont}
\rho_A[\sigma]=\frac{1}{Z}\int\mathcal D \alpha \exp{\left(-\pi^2 g^2_0\int dx\,dy\,\sigma^k(x)\Delta^{-1}(x-y)\sigma^k(y)+\frac{i}{2g^2_0}\int d^2 x\,\alpha(x)-\frac{N}{2}{\rm tr}\ln\Delta\right)},
\end{flalign}
\end{widetext}

\noindent with $\Delta(x)=-\nabla^2+i\alpha(x)$.

In order to make progress, we now can take the $N\rightarrow\infty$ limit, keeping $Ng_0^2$ fixed. In this limit, we can evaluate the integral
(\ref{Z_Cont}) by the method of steepest descent. The value of $\alpha$ that minimizes the action is given in the large $N$ limit by 
$\alpha(x)=-im^2$, with $m$ the solution of the equation

\begin{equation}\label{def_m}
 \int \frac{d^2k}{(2\pi)^2}\frac{1}{k^2+m^2}=\lim_{\Lambda\rightarrow\infty}\frac{1}{2\pi}\ln\left(\frac{\Lambda}{m}\right)=\frac{1}{Ng_0^2}.
\end{equation}

\noindent This equation is divergent, but can it can be made finite by renormalizing the bare coupling $g_0$ at an arbitrary renormalization
scale $M$ as $\frac{1}{g_0^2}=\frac{1}{g^2}+\frac{N}{2\pi}\ln\left(\frac{\Lambda}{M}\right).$ Inserting this equation back in (\ref{def_m}), 
we get the following expression for $m$ in terms of the physical coupling $g$, the renormalization scale $M$ and the number of components $N$ of
the original $\Omega$ field,

\begin{equation}
 m=M\exp\left[-\frac{2\pi}{g^2N}\right].
\end{equation}

In this large $N$ limit, we can compute the Entanglement Hamiltonian as the logarithm of the reduced density matrix, obtaining

\begin{flalign}\label{H_ent}
H_{\rm ent}=\ln\rho_A[\sigma]=(\pi g)^2\int dx\,dy\,\sigma^k(x)\Delta^{-1}(x-y)\sigma^k(y),
\end{flalign}

\noindent where $\Delta^{-1}(x)=K_0(m|x|)/2\pi$ is the zeroth order modified Bessel function. The exponential decay of $K_0(m|x|)$ for
large $x$ is what defines a local interaction at the boundary of $A$. Although this result is obtained in the large $N$ limit, the general
features of the $N=3$ model are believed to be captured in this limit \cite{Zamolodchikov}.

\section{Conclusions}\label{conclusions}

Given the structure of the VBS ground state, it is possible to define on any planar graph, without loops, a VBS state, where the local
spin at site $i$ is given by $z_i/2$, with $z_i$ the coordination number at site $i$. Using the Schwinger boson representation of the 
VBS ground state and the classical variable representation of this state, an expression for the partial density matrix $\rho_A$, which
describe the physical subsystem $A$, obtained by partitioning the whole unique ground state,
can be written. This expression for $\rho_A$ decompose into a classical loop expansion of the $O(3)$ model in the gapped phase.
Analyzing the different loop contributions, and assuming translation invariance, we have shown that the partial density matrix 
that describes a subsystem of the VBS ground state can be expressed as a sum over different
rotation-invariant quantum operators, where the Heisenberg interaction between nearest neighbors gives the largest nontrivial contribution
to the expansion. This quantum operators act on a spin $1/2$ chain in the boundary of the partition. The translational invariance assures us
that the different contributions along the boundary are equally weighted, so the boundary operator is given by the XXX Heisenberg Hamiltonian.
Here we discuss the case of translation invariant lattices which can be embedded in a torus, but for other lattices with different topologies
we expect similar results.

For non translational invariant lattices, the first nontrivial local interaction term is expected to be also of the type $\sigma_i\cdot\sigma_{i+1}$
but the Hamiltonian along the boundary will have different numerical prefactors for each local Heisenberg interactions, generating a non invariant Heisenberg Hamiltonian 
in the boundary.

In the continuous limit, we show that the entanglement Hamiltonian for this model is actually a local Hamiltonian, where the Hamiltonian density 
corresponds to a Heisenberg interaction of spin $1/2$ particles.

The analysis shown in this paper should be useful for studying other dimensions $d>2$ or other two dimensional lattices with more than one bond between
a pair of sites. In that case, the local dimension of the spin operators in the boundary Hamiltonian should increase, having then boundary chains
with higher representations of $SU(2)$ per site, but still with $SU(2)$ invariant local interactions.
 
\smallskip

\textbf{Acknowledgments} R. S. wish to thank F.N.C. Paraan for discussions and Dr. Tzu-Chieh Wei and Dr. Vladimir Korepin for careful reading this manuscript 
and for their useful comments. R.S particularly thanks Dr. Ignacio Cirac for discussions and the Max Planck Institute for Quantum Optics for 
hospitality. R.S is supported by a Fulbright-Conicyt Fellowship. 


\begin{thebibliography}{10}
\bibitem{com1}%
  \BibitemOpen
  \bibinfo {note} {We can write the operator $\rho_A$ as a matrix using the
  basis $|i\rangle=\sum_\alpha U_{i\alpha}|A_{\alpha}\rangle$ and its dual
  $\langle j|=\sum_\gamma V_{j\gamma}\langle A_{\gamma}|$, where $U$ and $V$
  are matrices of change of basis and $\langle j|i\rangle=\delta_{ij}$. We have
  $\langle j|\rho_A|i\rangle=(\rho_A)_{ji}={\mathcal
  N}\sum_{\alpha\beta\mu\gamma}\langle B_{\beta}|B_\alpha\rangle
  V_{j\gamma}\langle A_{\gamma}|A_\alpha\rangle U_{i\mu}\langle
  A_\beta|A_{\mu}\rangle$. However from the orthonormality condition $\langle
  j|i\rangle=\delta_{ij}$ follows the relation $\sum_\gamma V_{j\gamma}\langle
  A_{\gamma}|A_{\alpha}\rangle=(U^{-1})_{\alpha j}$ which inserted back in the
  expression for $(\rho_A)_{ji}$ simplifies it to $\langle
  j|\rho_A|i\rangle={\mathcal N}\sum_{\alpha\beta\mu}U_{i\mu}(\langle
  A_\mu|A_{\beta}\rangle)^*\langle B_{\beta}|B_\alpha\rangle(U^{-1})_{\alpha
  j}$,}%
  \bibAnnoteFile{NoStop}{com1}%
\end{thebibliography}

%

\end{document}